# Spacer-Mediated Gold Nanocube Arrays for Edge-Localized Excitonic Enhancement in Monolayer $MoS_2$

Abdullah Efe Yildiz[1,2] and Emre Ozan Polat[1,2,*]

[1]*Department of Physics, Bilkent University, 06800, Ankara, Turkey*

[2]*UNAM - National Nanotechnology Research Center and Institute of Materials Science and Nanotechnology, Bilkent University, Ankara 06800, Turkey*

**Corresponding author: emre.polat@bilkent.edu.tr*

## Abstract

Plasmonic nanostructures offer an effective route for enhancing light-matter interaction in atomically thin semiconductors, whose optical response is intrinsically limited by their sub-nanometer active thickness. Here, we numerically investigate excitonic enhancement in monolayer (ML) Molybdenum Disulfide ($MoS_2$) coupled to size-tuned gold (Au) nanocube arrays separated by thin aluminum oxide ($Al_2O_3$) and hexagonal boron nitride (h-BN) spacer layers. By varying the nanocube side length, the localized surface plasmon resonance is tuned across the visible spectral range to modulate the A- and B-excitonic transitions of monolayer $MoS_2$. We show that the nanocube-size-dependent spectral redshift can be further controlled through the spacer material and thickness, enabling systematic tuning of the near-field distribution, carrier generation rate, quantum yield, and radiative decay enhancement. Localized plasmonic confinement yields excitation-rate enhancements of up to 4.35 at B-excitonic transition (605 nm) and 3.66 at A-excitonic transition (650 nm), while the radiative decay-rate enhancement exceeds 80, leading to 350-fold photoluminescence enhancement. Although both A- and B-excitonic channels are enhanced simultaneously, their relative contributions depend on nanocube size, spacer material, and spacer thickness, indicating wavelength-dependent excitonic modulation rather than strict exciton-selective switching. These findings establish Au nanocube arrays as a simple, scalable, and tunable plasmonic platform for enhancing excitonic carrier generation and emission in ML $MoS_2$.

## Introduction

Two-dimensional (2D) transition-metal dichalcogenides (TMDs) are viable materials for nanoscale optoelectronics, photonics, and valleytronic device concepts due to their electronic band structure, which undergoes an indirect-to-direct gap transition [1,2]. This enables direct optical observation and manipulation of light-matter interactions and open avenues for new device applications that are not possible by wafer-based technologies [3-5]. Among TMDs, ML $MoS_2$ is one of the most widely investigated materials both experimentally and theoretically. The transition from an indirect band gap in bulk $MoS_2$ to a direct-gap semiconductor in its thinnest form produces photoluminescence (PL) and strong excitonic absorption features, making $MoS_2$ a convenient platform for studying light-matter interaction in atomically thin media. In addition, the broken inversion symmetry and strong spin-orbit coupling in group-VI TMD monolayers due to their large atomic weight give rise to coupled spin and valley degrees of freedom, which further increases the relevance of these materials for polarization-sensitive and valley-dependent photonic responses [6].

A central optical property of monolayer $MoS_2$ is the presence of pronounced A and B excitonic resonances [7]. These resonances originate from transitions near the K and K' valleys and are separated mainly by spin-orbit splitting of the valence band. Because dielectric screening is strongly reduced in the 2D limit, excitons in monolayer TMDs possess large oscillator strength and remain robust at room temperature compared to its bulk counterpart [8-11]. This strong excitonic character gives monolayer $MoS_2$ a much larger optical response than would normally be expected from an active layer with sub-nanometer thickness [12,13]. Nevertheless, the absolute interaction length remains extremely small around 0.7 nm, so efficient absorption, carrier generation, and radiative emission often require additional photonic or plasmonic strategies to concentrate the electromagnetic field near the ML [14,15].

To that end, plasmonic nanostructures provide a direct route to enhancing the optical response of atomically thin semiconductors. Localized surface plasmon resonances (LSPRs) in metallic nanoantennas can confine optical fields to subwavelength volumes, increase the local field intensity at the position of the semiconductor, and modify the local density of optical states (LDOS) that controls radiative decay [16-20]. In this picture, absorption enhancement is mainly associated with stronger near-field intensity at the monolayer, whereas emission enhancement depends not only on the excitation field but also on the balance between radiative and non-radiative decay channels [21-25]. Therefore, plasmonic nanoantenna design for monolayer $MoS_2$ should consider multiple parameters simultaneously, including spectral absorption, carrier generation rate, field enhancement, and radiative decay enhancement, rather than relying on a single electric field confinement metric.

Previous experimental and theoretical studies have shown that coupling monolayer $MoS_2$ to plasmonic nanostructures can enhance PL activity, reshape the excitonic spectrum, and modify radiative emission channels. Nanodisc arrays [26], nanorods [27], Fano-resonant antennas [28], nanoparticle dimers [29], and other antenna geometries [30] have been used to increase the electromagnetic field near $MoS_2$ and to improve exciton-plasmon coupling. These works demonstrate the general usefulness of plasmonic integration; however, they also show that the response is highly geometry-dependent and morphologic changes can affect the outcome of the optical activity drastically. The plasmon resonance position, linewidth, near-field distribution, radiative efficiency, and non-radiative loss channel all depend on nanoantenna shape, size, material, dielectric environment, and spacer thickness. As a result, a simple geometry that provides design flexibility for tunable plasmon resonances and strong near-field localization that are needed for nanophotonic device architecture to be experimentally implemented.

In this manner, Au nanocubes (AuNCs) are a particularly simple but useful geometry for plasmonic coupling because their flat facets and sharp corners support the in-depth analysis and control of strongly localized electromagnetic hot spots. Compared with round nanoparticles, such as discs and spheres, NCs provide stronger field concentration near edges and vertices, therefore their plasmonic response can be tuned easily by changing the cube side length and by modifying the surrounding dielectric environment [31-33]. Moreover, in array configurations, NC size can influence collective response, radiative scattering, and coupling to the far field. These properties make AuNCs promising for engineering exciton-plasmon interaction in two-dimensional semiconductors while maintaining a structure that is simple, easy to fabricate and compatible with colloidal, template-assisted, or lithographic fabrication routes.

Here, we numerically investigate 2D $MoS_2$ coupled to periodic AuNC arrays with variable side lengths. The AuNCs are separated from the $MoS_2$ monolayer by thin dielectric spacers, with $Al_2O_3$ and h-BN spacer materials with their variable thickness used as a key control parameter to observe the LSPR penetration. The optical response is analyzed through extinction spectra of AuNCs, absorption spectra of the hybrid $MoS_2$ structures, carrier generation rate, near-field intensity maps, and radiative decay enhancement. This complete set of design parameters allows us to correlate the plasmon resonance of the AuNCs to the excitonic A and B spectral region of $MoS_2$. With the provided in-depth analysis of the plasmon modulated excitonic spectra, we quantitatively report on the spectral reshaping, local-field enhancement, carrier generation, and emission-channels. The results show that size-tuned AuNCs provide a reliably tunable and experimentally applicable platform for spacer-mediated modulation of excitonic processes in monolayer $MoS_2$.

## Results

The geometry dependent plasmonic response of the AuNC system was first investigated before integration with monolayer $MoS_2$ to identify NC dimensions suitable for exciton-plasmon coupling. **Figure 1a** shows the AuNC geometry with side length *a*, that is placed above a dielectric spacer of thickness ($t_s$). NC structure is optically effective due to flat facets, and sharp cube corners supporting localized electromagnetic hot spots while simultaneously allowing the incidence of electromagnetic wave at the top of the structure. Such field localization is especially useful for atomically thin semiconductors, where the optical response is immensely sensitive to the near-field intensity sampled by the monolayer.

Among the set of side lengths, we found exciton-peak-matching geometries and implemented them on the ML $MoS_2$**. Figures 1b** and **1c** show the near-field intensity distributions for the 105 and 130 nm wide AuNCs, that serves for the exciton modulation purpose respectively. These two AuNC sizes produce varying enhancement strengths, showing that the side length modifies not only the spectral intensity of the plasmon resonance but also the spatial profile of the near field. This is crucial for the underlying excitonic system because the monolayer interacts only with the field that reaches the spacer/ ML $MoS_2$ interface.

The size-dependent extinction spectra of the AuNC are shown in **Figure 1d**. As the side length increases from 50 to 130 nm, the extinction intensity increases, because of the size and the main localized surface plasmon resonance gradually redshifts, however the shift is not as strong as the previous studies that use different shapes [34]. This trend is expected because larger metallic nanostructures allow longer effective charge oscillation lengths, shifting the plasmonic response toward lower energies. Among the investigated NC sizes, the 105 and 130 nm NCs provide the most relevant spectral overlap with the A- and B-excitonic region of monolayer $MoS_2$.

We further analyze the spacer media to pinpoint the overall excitonic contributions**. Figures 1e** and **1f** compare the extinction spectra of the 105 and 130 nm AuNCs on $Al_2O_3$ and h-BN dielectric spacers. Both spacer materials modify the plasmonic response by changing the local dielectric environment around the AuNC. For the 105 nm long NC, the LSPR peak shifts in different rates nm with for $Al_2O_3$ (from 589.2 to 609.3) and nm with h-BN (from 590.8 to 619.8) as the spacer thickness increases from 2 to 10 nm. Likewise, the 130 nm wide AuNC shows pronounced LSPR peak shifts from 611.0 to 633.6 nm with $Al_2O_3$ and from 610.1 to 648.9 nm with h-BN spacer. As summarized in **Table 1**, these values show that the plasmonic resonance can be tuned through both cube size and spacer environment, noting that the 130 nm wide AuNC gives a more redshifted response, while the 105 nm wide AuNC remains closer to the higher-energy part of the excitonic region.

The spacer-dependent spectral shift can be understood in terms of dielectric screening of the LSPR. As the spacer thickness increases, the evanescent near field of the AuNC samples a larger dielectric volume between the metal and the underlying $MoS_2/SiO_2/Si$ stack. This changes the effective optical environment experienced by the plasmonic mode and produces an observable redshift of the extinction spectrum. The magnitude of this shift depends on the spacer material because $Al_2O_3$ and h-BN modify the local dielectric screening differently.

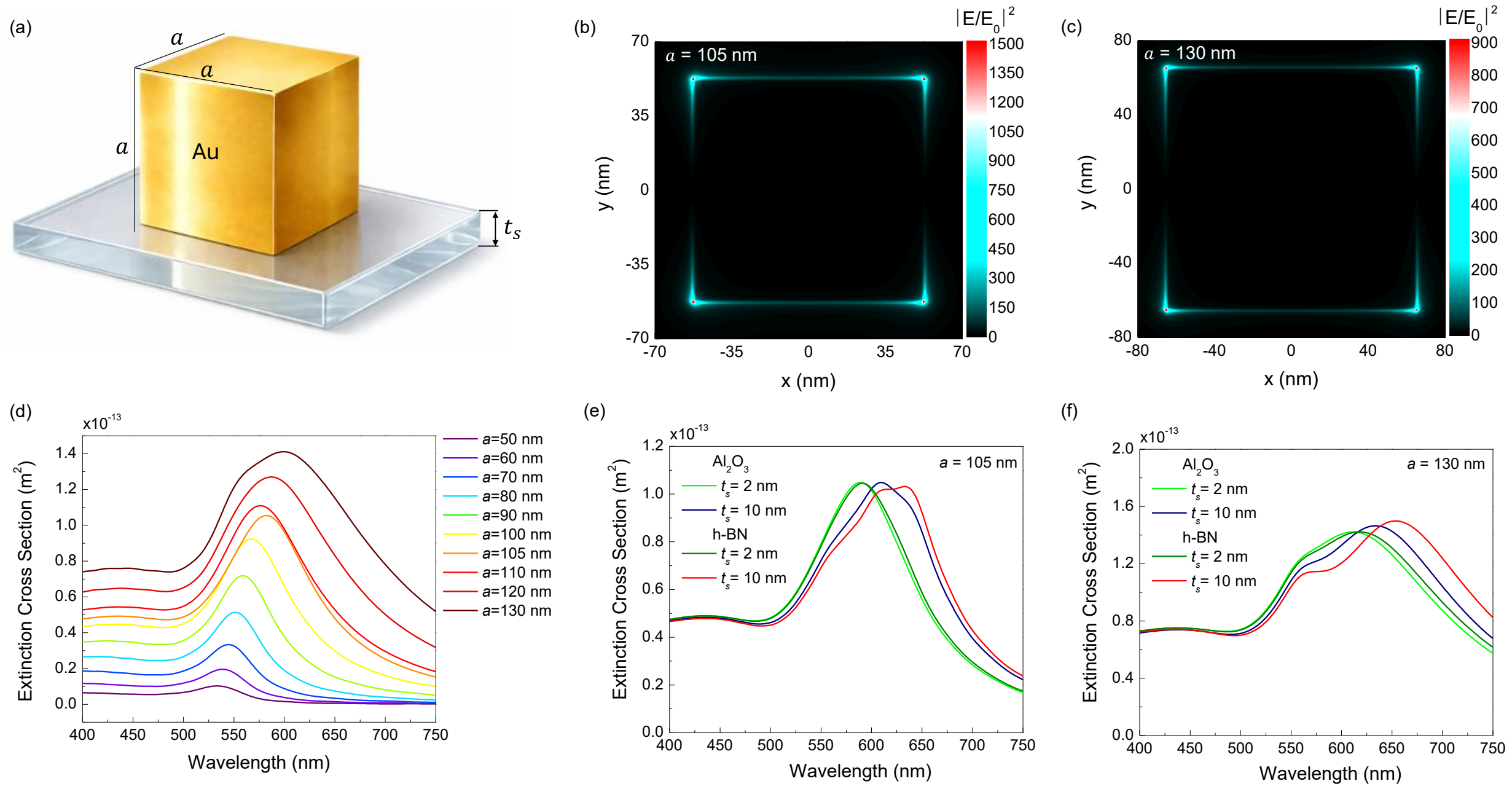

***Figure 1. Spacer mediated plasmonic response and design parameters of AuNCs.*** ***(a)*** *Schematic illustration of a single AuNC with side length a placed on a dielectric spacer of thickness* $t_s$. ***(b,c)*** *Near-field intensity maps of AuNCs with side lengths of 105 and 130 nm, respectively. Showing strong electric-field localization at the cube corners.* ***(d)*** *Extinction cross-section spectra for AuNCs with different side lengths from 50 to 130 nm, demonstrating the increase and redshift of the localized surface plasmon resonance with increasing cube size.* ***(e,f)*** *Extinction spectra of the 105 and 130 nm AuNCs in the presence of* $Al_2O_3$ *and h-BN spacer environments with different spacer thicknesses. The spectra indicate that both AuNC size and dielectric environment control the spectral overlap between the plasmonic mode and the* $MoS_2$ *excitonic region.*

This behavior is consistent with a simplified sensing-volume model, where the plasmon resonance shift depends on the effective refractive index sampled within the evanescent near-field decay length [35, 36]:

$$\Delta\lambda = m\Delta n\left[1 - e^{-2t_s/l_d}\right] \quad (1)$$

where ($t_s$) is the spacer thickness, ($l_d$) is the evanescent near-field decay length, ($m$) is the bulk refractive-index sensitivity of the plasmon mode, and ($\Delta n$) represents the effective refractive-index contrast introduced by the spacer. The exponential term describes the finite spatial extent of the plasmonic near field. At small spacer thicknesses, the NC mode is strongly influenced by the dielectric environment close to the metal/spacer interface. At larger thickness, the additional dielectric volume lies farther from the strongest near-field region, so the resonance shift gradually becomes less sensitive to further spacer increase. Unlike particle-on-mirror nanocavities [37], where the spacer mainly controls metal-metal gap-mode hybridization of the LSPR mode, the present dielectric-supported AuNC system is governed by near-field sampling of the spacer and substrate environment. The flat facet on top of the spacer and sharp bottom corners of the AuNC increases the overlap between the localized plasmonic field and the spacer layer. Therefore, changing ($t_s$) modifies the effective dielectric screening around the plasmonic mode and tunes the resonance position without changing the NC material or the overall periodic structure.

***Table 1. LSPR of AuNCs integrated with two different spacer materials.*** *The table lists the LSPR peak wavelengths and the corresponding spectral shifts relative to the bare AuNC for two side lengths, 105 and 130 nm. $Al_2O_3$ and h-BN spacer layers with thicknesses from 2 to 10 nm are compared to evaluate how dielectric environment and metal-$MoS_2$ separation tune the spectral overlap between the plasmonic resonance and the A- and B-excitonic transitions of ML $MoS_2$.*

| $a = 130\ nm$ | LSPR Peak, $\lambda$ (nm) | | Peak Shift, $\Delta\lambda$ (nm) | |
|---|---|---|---|---|
| Spacer Thickness (nm) | $Al_2O_3$ | h-BN | $Al_2O_3$ | h-BN |
| 2 | 611.0 | 610.1 | 12.0 | 11.1 |
| 4 | 618.9 | 627.1 | 19.9 | 28.1 |
| 6 | 624.3 | 637.4 | 25.3 | 38.3 |
| 8 | 628.9 | 644.1 | 29.9 | 45.0 |
| 10 | 633.6 | 648.9 | 34.5 | 49.9 |
| $a = 105\ nm$ | LSPR Peak, $\lambda$ (nm) | | Peak Shift, $\Delta\lambda$ (nm) | |
| Spacer Thickness (nm) | $Al_2O_3$ | h-BN | $Al_2O_3$ | h-BN |
| 2 | 589.2 | 590.8 | 7.2 | 8.8 |
| 4 | 595.7 | 598.2 | 13.7 | 16.3 |
| 6 | 600.7 | 604.1 | 18.8 | 22.2 |
| 8 | 605.8 | 609.3 | 23.9 | 27.3 |
| 10 | 609.3 | 619.8 | 27.3 | 37.8 |

**Figure 2a** demonstrates the integrated AuNC together with spacer material on ML $MoS_2$ hosting substrate ($SiO_2$/Si). **Figures 2b** and **2c** show the wavelength- and spacer-thickness-dependent normalized in-plane E-field intensity, evaluated at the $MoS_2$ plane for the 105 and 130 nm AuNCs with $Al_2O_3$ spacer. For both AuNC sizes, the strongest field enhancement occurs at small $t_s$ and decreases as $t_s$ increases. The 105 nm cube provides a field-enhancement region closer to the higher energy side of the excitonic window, whereas the 130 nm cube shows a more redshifted response. This confirms that the side length controls the spectral position of the plasmonic near-field enhancement, while the spacer thickness controls how strongly this enhancement reaches the $MoS_2$ monolayer.

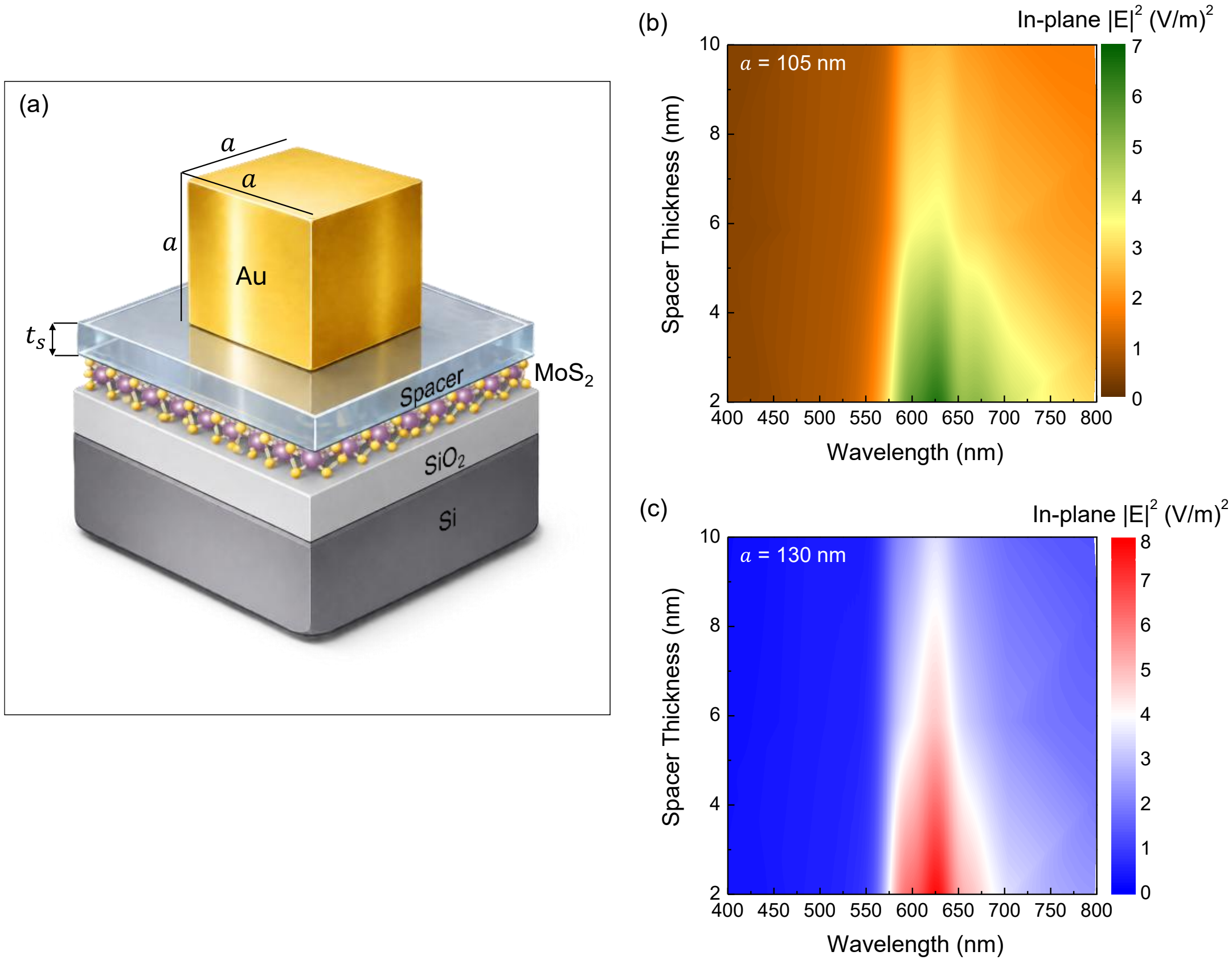

***Figure 2. Hybrid AuNC/spacer/MoS₂ device geometry yielding in-plane electric-field enhancement****. **(a)** Schematic of the AuNC/spacer/ML $MoS_2$ on $SiO_2$/Si device substrate. Modeled plasmonic device focuses the in plane electric field intensity for the enhanced exciton generation in $MoS_2$ layer. **(b)** Spectral in-plane electric field intensity map ($|E|^2$) for varying $Al_2O_3$ spacer thickness with 105 wide and **(c)** 130 nm wide AuNCs. The field enhancement is strongest near the plasmon-exciton overlap region and decreases or redistributes as the spacer thickness increases.*

The strongest enhancement is again obtained for the thinnest spacer layers, where the $MoS_2$ monolayer remains closest to the AuNC hot-spot region. As the $t_s$ increases, the field intensity at the $MoS_2$ plane gradually decreases because of the evanescent decay of the plasmonic near field. Overall, the side length $a$ controls the spectral position of the near-field enhancement, while the spacer thickness controls how efficiently this enhanced field reaches the monolayer $MoS_2$.

**Figure 3** presents the normalized absorption spectra of monolayer $MoS_2$ coupled to spacer-mediated AuNC structures and compares them with the bare $MoS_2/SiO_2/Si$ as the baseline. As shown in **Figure 3a**, the AuNCs are arranged periodically above the spacer/$MoS_2/SiO_2/Si$ stack, allowing the plasmonic near field generated by the cube corners to interact with the monolayer. The vertical dashed lines indicate B- and A-exciton positions of monolayer $MoS_2$. In all cases, the characteristic excitonic features of $MoS_2$ remain visible after integration with the AuNC and dielectric spacer. This shows that the intrinsic excitonic response of the monolayer is preserved in the hybrid structure. Rather than producing a large displacement of the exciton peaks, the AuNC mainly modifies the spectral background and changes the relative absorption intensity around the excitonic region.

For the 105 nm AuNC system shown in **Figure 3b**, the absorption spectra obtained with $Al_2O_3$ and h-BN spacers are relatively close to each other. Only moderate changes are observed around the A- and B-exciton wavelengths, indicating that this structure does not produce strong absorption selectivity between the two excitonic transitions. The main effect is a weak plasmon-assisted reshaping of the normalized absorption profile. The absence of clear peak splitting or large exciton shifts suggests that the system remains in a weak-coupling regime, where the AuNC modifies the local optical field without strongly changing the excitonic energy structure of $MoS_2$. **Figures 3c** and **3d** show the absorption spectra of the 130 nm AuNC with h-BN and $Al_2O_3$ spacers, respectively. Similar to the 105 nm case, the excitonic double-feature in the A/B region is retained for all $t_s$. Increasing the spacer thickness from 2 to 10 nm produces only small changes in the normalized absorption spectra. This weak thickness dependence is reasonable because the plotted spectra are normalized and since the total absorption response contains contributions from both the $MoS_2$ excitons and the broader plasmonic background of the AuNC.

AuNC integration preserves the excitonic absorption character of monolayer $MoS_2$ while introducing moderate size- and spacer-dependent spectral modulation. The absorption spectra alone do not provide strong evidence for A- or B-exciton-selective enhancement. Instead, they show that the plasmonic structure modifies the optical environment of $MoS_2$ without suppressing its intrinsic excitonic features. Therefore, the more direct evidence for wavelength-dependent enhancement is obtained from the carrier generation rate and radiative decay calculations, where excitation and emission-channel modifications are quantified more clearly.

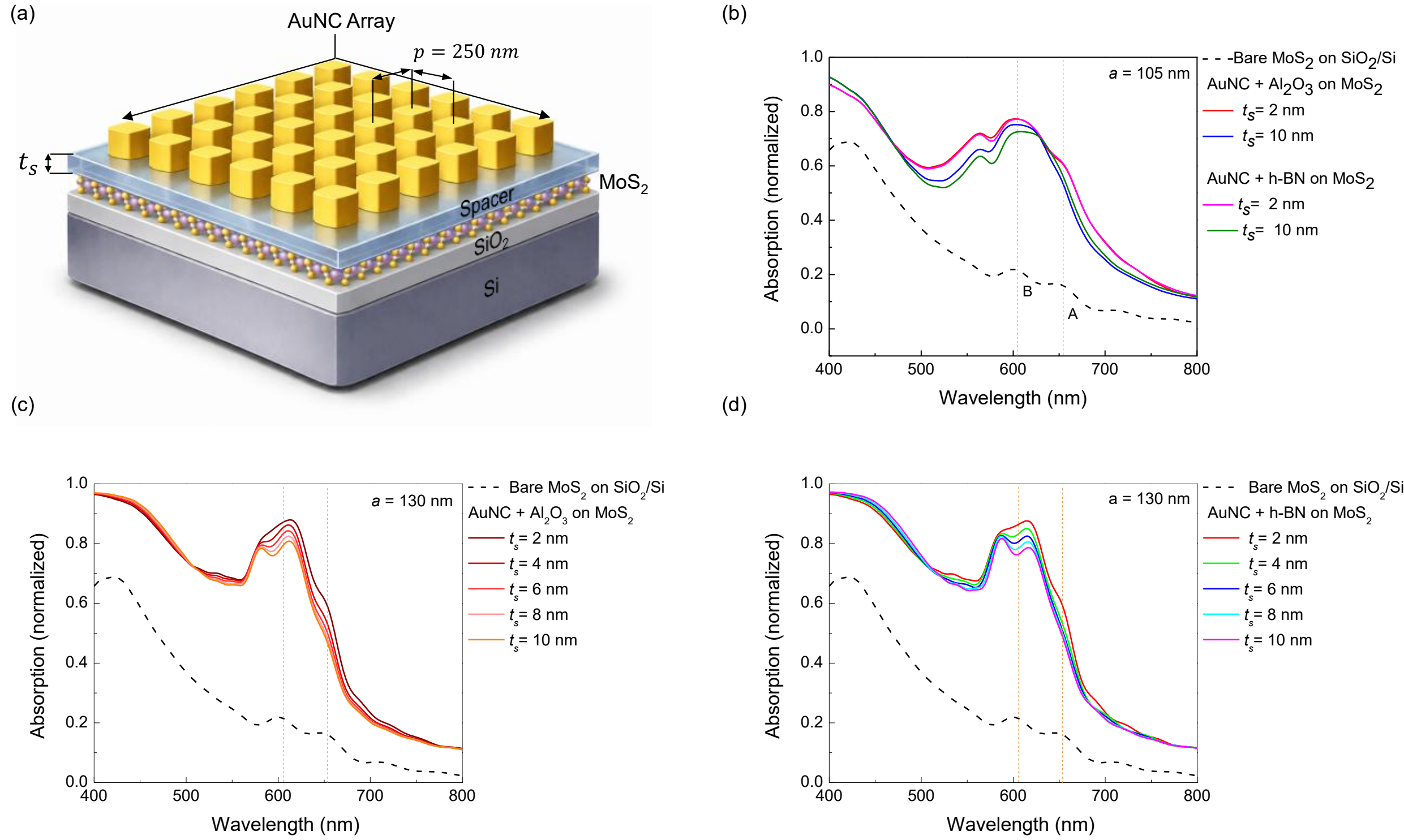

***Figure 3. Absorption spectra of monolayer $MoS_2$ coupled to AuNC structures through $Al_2O_3$ and h-BN spacers.*** *(a) Schematic of the periodic* AuNC *array integrated with the spacer/$MoS_2$/$SiO_2$/Si stack. The array period is $p = 250$ nm. (b) AuNC Normalized absorption spectra for $MoS_2$ coupled to 105 nm wide AuNC through different spacers. The dashed line shows the excitons of bare monolayer $MoS_2$ on $SiO_2$/Si. (c,d) Normalized absorption spectra for 130 nm AuNC structures with h-BN and $Al_2O_3$ spacers, respectively, for different $t_s$. The vertical dashed lines mark the B and A exciton wavelengths of monolayer $MoS_2$. The AuNC modifies the excitonic absorption profile through plasmon-assisted near-field coupling rather than producing a uniform broadband increase.*

**Figure 4** presents the carrier generation rate of monolayer $MoS_2$ coupled to periodic AuNC arrays. Unlike the normalized absorption spectra in **Figure 3**, the carrier generation rate gives a more direct measure of how efficiently incident light produces electron-hole pairs inside the $MoS_2$ monolayer. Therefore, this analysis is useful for evaluating the practical effect of plasmonic near-field enhancement on excitonic light harvesting. **Figures 4a** and **4b** show the carrier generation rate spectra for 130 nm AuNC with h-BN and $Al_2O_3$ spacers, respectively. In both spacer systems, the carrier generation is strongly enhanced in the visible excitonic region of $MoS_2$, especially near the A- and B-exciton wavelengths. For the 105 nm AuNC counterpart shown in **Figures 4c** and **d**, the carrier generation spectra follow the same spacer-dependent trend, but with a slightly lower overall magnitude compared with the 130 nm structures. This indicates that the smaller nanocube still enhances excitonic carrier generation efficiently, while the reduced cube size shifts and weakens the plasmonic overlap with the $MoS_2$ excitonic region relative to the 130 nm AuNC. The enhancement is largest for the thinnest spacer layers and gradually decreases as the spacer thickness increases from 2 to 10 nm. This trend agrees with the electric-field maps in **Figure 2**, confirming that carrier generation is mainly controlled by the near-field intensity reaching the $MoS_2$ plane. When the $t_s$ is small, $MoS_2$ is located close to the AuNC hot spots; when the $t_s$ get thicker, the evanescent field decays before reaching the monolayer, reducing the generation rate.

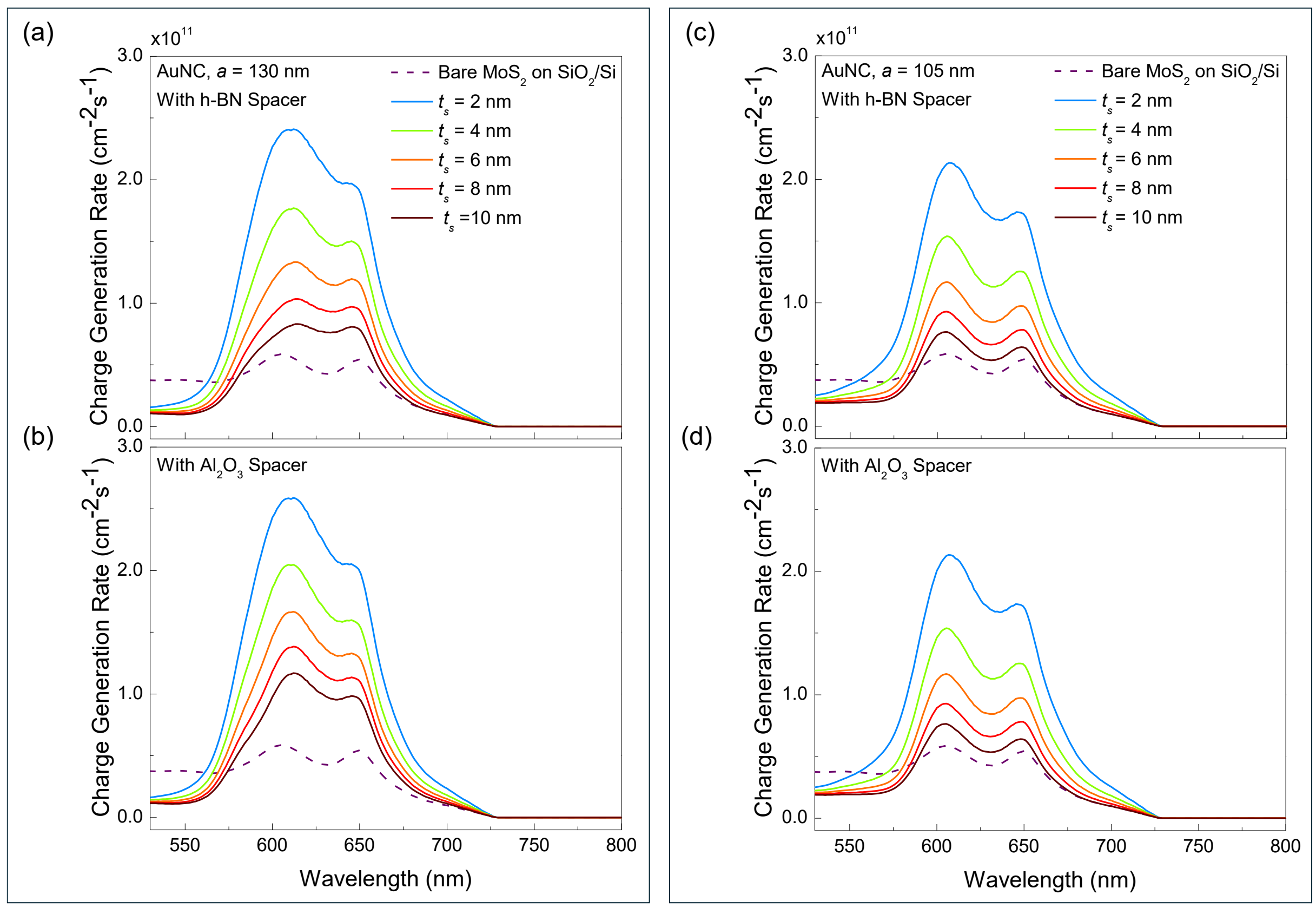

***Figure 4. Carrier generation rate spectra of ML $MoS_2$ coupled to AuNC arrays for different spacer materials and thicknesses.*** ***(a,b)*** *The carrier generation spectra for the 130 nm AuNC integrated with h-BN and $Al_2O_3$ on ML $MoS_2$ respectively.* ***(c,d)*** *The corresponding spectral charge generation rates for the 105 nm AuNC. In all cases, the generation rate is enhanced around the A- and B-excitonic peaks of monolayer $MoS_2$, with the strongest response obtained for the thinnest spacer.*

The quantitative excitation enhancement factors are summarized in **Table 2**. For the 130 nm AuNC array on a 2 nm $Al_2O_3$ spacer, the excitation enhancement reaches 3.66 at the A-exciton wavelength of 650 nm and 4.35 at the B-exciton wavelength of 605 nm. As the spacer thickness increases to 10 nm, these values decrease to 1.76 and 1.87, respectively, reflecting the reduced near-field coupling between the AuNC array and the monolayer $MoS_2$. A similar trend is observed for the h-BN spacer, where the enhancement factors are 3.50 and 4.05 for the A- and B-excitonic transitions at 2 nm spacer thickness, decreasing to 1.44 and 1.32 at 10 nm.

***Table 2. Excitation enhancement, radiative decay-rate enhancement, and photoluminescence enhancement of monolayer $MoS_2$ coupled to AuNCs.*** *The values are compared for $Al_2O_3$ and h-BN spacers with thicknesses from 2 to 10 nm. Each entry is reported for A and B excitonic transitions at 650 nm and 605 nm respectively. The results show that thinner spacers provide stronger enhancement, while the spacer material and AuNC size modify the relative response at both excitonic wavelengths.*

| **Side length, *a= 130 nm*** | **Excitation enhancement factor** $(\frac{G}{G_0})$ | | | | **Normalized radiative decay rate** $(F_{rad})$ | | | | **PL enhancement** $(PLE)$ | | | |
|---|---|---|---|---|---|---|---|---|---|---|---|---|
| **Spacer thickness, $t_s$ (nm)** | **$Al_2O_3$** | | **h-BN** | | **$Al_2O_3$** | | **h-BN** | | **$Al_2O_3$** | | **h-BN** | |
| | **A exciton (@650 nm)** | **B exciton (@605 nm)** | **A exciton (@650 nm)** | **B exciton (@605 nm)** | **A exciton (@650 nm)** | **B exciton (@605 nm)** | **A exciton (@650 nm)** | **B exciton (@605 nm)** | **A exciton (@650 nm)** | **B exciton (@605 nm)** | **A exciton (@650 nm)** | **B exciton (@605 nm)** |
| 2 | 3.66 | 4.35 | 3.50 | 4.05 | 49.30 | 81.30 | 50.55 | 70.17 | 180.44 | 353.66 | 176.93 | 284.19 |
| 4 | 2.85 | 3.41 | 2.66 | 2.92 | 29.53 | 41.68 | 25.24 | 29.34 | 84.16 | 142.13 | 67.14 | 85.67 |
| 6 | 2.37 | 2.74 | 2.13 | 2.17 | 19.31 | 23.44 | 14.86 | 14.57 | 45.76 | 64.23 | 31.65 | 31.62 |
| 8 | 2.03 | 2.24 | 1.73 | 1.66 | 13.46 | 14.44 | 9.67 | 8.27 | 27.32 | 32.35 | 16.73 | 13.73 |
| 10 | 1.76 | 1.87 | 1.44 | 1.32 | 9.81 | 9.51 | 6.71 | 5.18 | 17.27 | 17.78 | 9.66 | 6.84 |
| **Side length, *a= 105 nm*** | **Excitation enhancement factor** $(\frac{G}{G_0})$ | | | | **Normalized radiative decay rate** $(F_{rad})$ | | | | **PL enhancement** **(*PLE*)** | | | |
| **Spacer thickness, $t_s$ (nm)** | **$Al_2O_3$** | | **h-BN** | | **$Al_2O_3$** | | **h-BN** | | **$Al_2O_3$** | | **h-BN** | |
| | **A exciton (@650 nm)** | **B exciton (@605 nm)** | **A exciton (@650 nm)** | **B exciton (@605 nm)** | **A exciton (@650 nm)** | **B exciton (@605 nm)** | **A exciton (@650 nm)** | **B exciton (@605 nm)** | **A exciton (@650 nm)** | **B exciton (@605 nm)** | **A exciton (@650 nm)** | **B exciton (@605 nm)** |
| 2 | 3.36 | 3.88 | 3.14 | 3.61 | 49.44 | 91.72 | 46.86 | 70.62 | 166.12 | 355.87 | 147.14 | 254.94 |
| 4 | 2.59 | 2.98 | 2.28 | 2.62 | 32.23 | 49.02 | 26.01 | 30.76 | 83.48 | 146.08 | 59.30 | 80.59 |
| 6 | 2.11 | 2.35 | 1.78 | 1.99 | 21.75 | 27.51 | 15.74 | 15.24 | 45.89 | 64.65 | 28.02 | 30.33 |
| 8 | 1.77 | 1.93 | 1.43 | 1.58 | 15.47 | 16.76 | 10.27 | 8.58 | 27.38 | 32.35 | 14.69 | 13.56 |
| 10 | 1.51 | 1.62 | 1.17 | 1.30 | 11.43 | 10.87 | 7.06 | 5.37 | 17.26 | 17.61 | 8.26 | 6.98 |

For the 105 nm AuNC array, the excitation enhancement remains significant but is generally weaker than that of the 130 nm array. In the case of the 2 nm $Al_2O_3$ spacer, the enhancement reaches 3.36 for the A exciton and 3.88 for the B exciton, while the corresponding values for the h-BN spacer are 3.14 and 3.61. Overall, these results show that both NC geometries enhance excitonic carrier generation in monolayer $MoS_2$, with the 130 nm AuNC array providing stronger enhancement due to its more favorable spectral overlap with the excitonic transitions. The enhancement is typically more pronounced near the B-excitonic transition; however, both A- and B-excitonic channels are enhanced simultaneously, indicating wavelength-dependent excitonic modulation rather than strict exciton-selective enhancement.

For materials with intrinsically low quantum yield such as monolayer $MoS_2$ ($\eta_0 \approx 10^{-3}$-$10^{-4}$), the *PLE* can be approximated as: [30,38]

$$PLE \approx \left|\frac{G}{G_0}\right| \times F_{rad} \tag{2}$$

where $G/G_0$ represents the *CGR* (excitation) enhancement factor and $F_{rad}$ is the normalized radiative decay rate of the hybrid system. The radiative and non-radiative decay rates are calculated using an in-plane dipole model for the $MoS_2$ exciton interacting with the plasmonic nanocavity, allowing the quantum efficiency of the hybrid structure to be expressed as:

$$QE = \frac{F_{rad}}{F_{rad}+F_{nonrad}} \tag{3}$$

We evaluate the resulting quantum yield, radiative decay-rate enhancement, and photoluminescence enhancement of the AuNC-$MoS_2$ hybrid system in **Figure 5**. **Figure 5a** present the quantum-yield map for the 105 nm AuNC with $Al_2O_3$ spacer . The quantum yield shows a clear dependence on both wavelength and spacer thickness, reflecting the balance between enhanced radiative emission and non-radiative energy transfer to the metal. At small spacer thicknesses, the $MoS_2$ monolayer strongly couples to the plasmonic near field of the AuNC, but metal-induced losses can also become significant. As the spacer thickness increases, non-radiative loss is reduced, although the overall plasmonic coupling strength also weakens. Therefore, the quantum-yield maps highlight the need to optimize the metal-$MoS_2$ separation rather than simply minimizing the spacer thickness.

**Figure 5b** depicts the normalized radiative decay-rate enhancement at the A- and B-excitonic spectral regions with the 105 and 130 nm AuNC using $Al_2O_3$ spacer. The wavelength dependence of the radiative enhancement reveals that the response is not spectrally uniform. In most configurations, the enhancement at 605 nm is larger than or comparable to that at 650 nm, showing that the B-exciton-side response benefits strongly from the AuNC near field. For instance, the 105 nm AuNC with $Al_2O_3$ spacer gives radiative enhancement values of 49.44 at 650 and 91.72 at 605 nm for the 2 nm spacer, while the 130 nm AuNC with $Al_2O_3$ gives 49.30 and 81.30 at the same A and B excitonic wavelengths. This behavior further supports that the AuNC geometry provides wavelength-dependent emission enhancement, with a slight preference toward the B-exciton (605 nm) rather than equal enhancement across both excitonic channels.

**Figure 5c and 5d** present the recorded spectral quantum yield and normalized radiative decay-rate enhancement for the h-BN spacer counterpart. These results are directly connected to the carrier generation enhancement shown in **Figure 4**. While **Figure 4** demonstrates that thinner spacers provide stronger excitation enhancement, **Figure 5** shows that the radiative emission channel follows a similar distance-dependent trend. The resulting photoluminescence enhancement factors are also summarized in **Table 2**. For the 130 nm AuNC array with a 2 nm $Al_2O_3$ spacer, the estimated photoluminescence enhancement reaches 180.44 at the A-exciton wavelength of 650 nm and 353.66 at the B-exciton wavelength of 605 nm. For the same spacer thickness, the corresponding values with h-BN are 176.93 and 284.19, respectively. On the other hand, using 105 nm AuNC array provides the photoluminescence enhancement factors such as 166.12 and 355.87 for $Al_2O_3$, and 147.14 and 254.94 for h-BN at 650 and 605 nm, respectively. These large enhancement values arise from the combined increase in the excitation rate and radiative decay channel. As the spacer thickness increases, the photoluminescence enhancement decreases substantially, following the reduction in both carrier generation and radiative decay enhancement. This behavior confirms that the strongest overall response is obtained when monolayer $MoS_2$ is positioned sufficiently close to the AuNC hot spots to benefit from near-field coupling. Therefore, the photoluminescence enhancement should be interpreted as the combined outcome of spectral plasmon-exciton overlap and distance-dependent near-field coupling, rather than as a simple consequence of NC size alone.

Overall, **Figure 5** and **Table 2** demonstrate that the AuNC geometry provides strong plasmon-assisted emission enhancement through corner-localized hot spots, while the spacer material and thickness control the balance between radiative enhancement and metal-induced loss. The results also show that the optimum photoluminescence response is not governed by absorption enhancement alone, but by the combined behavior of excitation enhancement, emission efficiency, and radiative decay-rate enhancement. In most configurations, $Al_2O_3$ provides slightly stronger overall performance than h-BN, particularly for the thinnest spacer layers, yielding higher excitation enhancement, radiative decay enhancement, and photoluminescence enhancement factors. Therefore, the AuNC-$MoS_2$ hybrid system offers a simple and tunable route for engineering excitonic photoluminescence through NC size, spacer material, and metal-semiconductor separation. We emphasize that, although the AuNC structures strongly enhance the excitonic emission response, the results do not indicate strict A- or B-exciton selectivity. Both the 605 and 650 nm channels are enhanced simultaneously, and their relative enhancement varies with NC size, spacer material, and spacer thickness. Thus, the present system is more accurately described as enabling wavelength-dependent excitonic enhancement rather than complete exciton-selective switching. This distinction is important, as the NC geometry enhances both excitation and emission channels while preserving the intrinsic excitonic features of monolayer $MoS_2$.

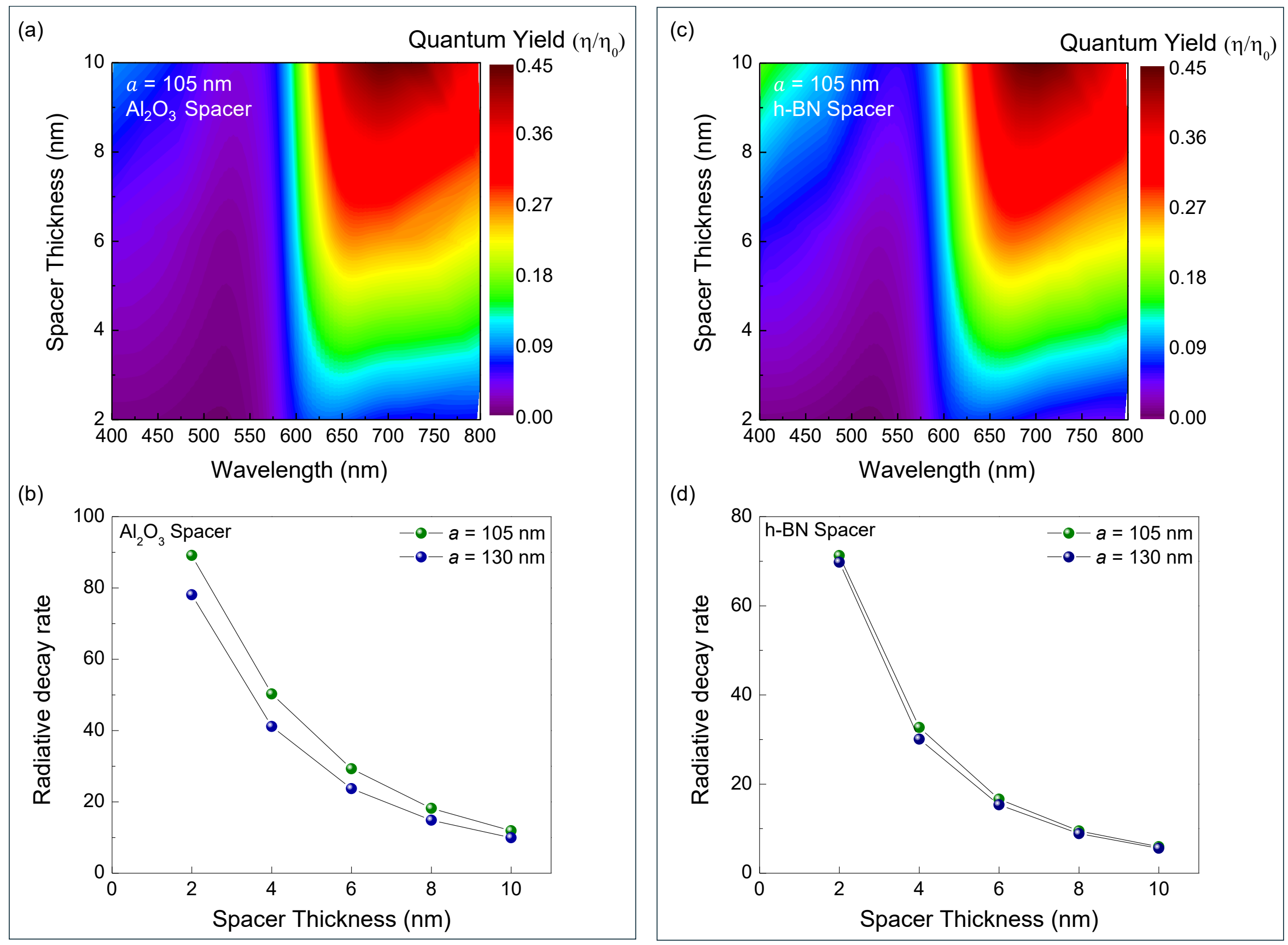

***Figure 5. Radiative decay enhancement and resulting quantum yield by the control of spacer thickness and AuNC size.*** ***(a)*** *Wavelength- and spacer-thickness-dependent quantum yield response map for the 105 nm* AuNC *with $Al_2O_3$ spacer.* ***(b)*** *Radiative decay enhancement as a function of spacer thickness for 105 and 130 nm AuNCs with $Al_2O_3$ spacer.* ***(c)*** *Wavelength- and spacer-thickness-dependent quantum yield response map for the 105 nm* AuNC *with h-BN spacer.* ***(d)*** *Radiative decay enhancement as a function of spacer thickness for 105 and 130 nm AuNCs with h-BN spacer. The enhancement is strongest for the thinnest spacer and decreases rapidly with increasing spacer thickness, reflecting the evanescent distance dependence of plasmon-exciton coupling*

## Conclusion, Discussion and Outlook

In this work, we numerically investigated the plasmonic modulation of excitonic optical processes in monolayer $MoS_2$ using size-tuned periodic AuNC arrays. Two NC side lengths, 105 and 130 nm, were studied together with $Al_2O_3$ and h-BN spacer layers. The results show that AuNCs provide a simple and effective plasmonic platform because their sharp corners support localized electromagnetic hot spots, while their side length controls the spectral position of the localized surface plasmon resonance.

The extinction spectra show that increasing the AuNC size redshifts and strengthens the plasmonic response, allowing improved overlap with the A- and B-excitonic region of monolayer $MoS_2$. Near-field maps further confirm that the strongest field localization occurs near the cube corners and at small spacer thicknesses, where the $MoS_2$ monolayer is positioned close to the evanescent plasmonic near field. This enhanced local field modifies the excitonic optical response and increases the carrier generation rate around the intrinsic excitonic transitions of $MoS_2$.

The spacer layer was used as a key design parameter. Thinner spacers provide stronger near-field coupling and therefore larger carrier generation and radiative decay-rate enhancement, while thicker spacers reduce the interaction because of the rapid decay of the plasmonic field. The comparison between $Al_2O_3$ and h-BN also shows that the spacer material changes the dielectric environment around the AuNC, modifying the resonance position, field distribution, and emission response. As a result, the final photoluminescence enhancement depends on the combined effect of AuNC size, spacer material, and spacer thickness.

Although the AuNC structures enhance both the 605 and 650 nm excitonic regions, the system does not demonstrate strict A- or B-exciton selectivity. Instead, it provides wavelength-dependent excitonic enhancement, where both excitonic channels are improved but with different relative strengths depending on the geometry and spacer configuration. This distinction is important because the NC platform should be described as a compact plasmonic enhancer rather than a complete exciton-selective switching system.

Overall, these results establish AuNC arrays as a physically simple and experimentally accessible design for enhancing excitonic processes in atomically thin $MoS_2$. The combined control of AuNC size, spacer thickness, and spacer material enables tunable exciton-plasmon coupling without requiring external fields, chemical doping, or complex active tuning schemes. Such structures may be useful for enhanced 2D-material photodetectors, nanoscale light emitters, excitonic sensors, and integrated nanophotonic devices based on transition-metal dichalcogenide monolayers.

## Methodology

To accurately capture both the atomistic excitonic response of monolayer $MoS_2$ and the electromagnetic field confinement of the plasmonic cavity, we combined first-principles excitonic calculations with full-wave electromagnetic simulations. The optical response of the $MoS_2$-AuNC system was simulated using the FDTD method, which numerically solves Maxwell's equations in both space and time. In this approach, Maxwell's curl equations

$$\nabla \times \mathbf{E} = -\mu \frac{\partial \mathbf{H}}{\partial t} \tag{4}$$

$$\nabla \times \mathbf{H} = \varepsilon \frac{\partial \mathbf{E}}{\partial t} + \mathbf{J} \tag{5}$$

are discretized on a Yee grid, where **E** and **H** denote the electric and magnetic fields, μ is the magnetic permeability, ε is the permittivity, and **J** is the induced current density. Perfectly matched layers (PMLs) were applied at the simulation boundaries to emulate an open electromagnetic environment and suppress artificial reflections. All simulations were performed using a three-dimensional FDTD solver (Ansys Lumerical FDTD) with spatial and temporal discretization chosen to satisfy the Courant stability condition while resolving the nanoscale features of the $MoS_2$ layer and the plasmonic cavity.

The dispersive optical properties of the materials were implemented using experimentally validated refractive-index datasets. Gold was described using the Johnson and Christy [39] data set, while Si, $SiO_2$, and $Al_2O_3$ were modeled using wavelength-dependent optical constants from the Palik Handbook [40-42]. Meanwhile, h-BN was implemented as a dielectric using its ordinary and extraordinary dispersion relation [43].

To evaluate the optical response of the plasmonic cavity, broadband extinction spectra were calculated using a total-field scattered-field (TFSF) plane-wave source spanning 400-800 nm. Absorption within the AuNC and spacer layers was obtained by integrating the power dissipation density over their volumes, while scattering was calculated from the far-field power exiting the TFSF boundaries.

Optical constants of monolayer $MoS_2$ is taken experimentally [44] and the resulting complex dielectric function:

$$\varepsilon(\omega) = \varepsilon_1(\omega) + i\varepsilon_2(\omega) \tag{6}$$

was implemented in the FDTD simulations as a surface conductivity layer according to:

$$\sigma(\omega) = -i\omega t \varepsilon_0 [\varepsilon(\omega) - 1] \tag{7}$$

where $\varepsilon_0$ is the vacuum permittivity, $\omega$ is the angular frequency, and $t$ is the effective monolayer thickness. The $MoS_2$ layer was modeled as an in-plane isotropic sheet conductivity, consistent with the dominant in-plane excitonic response of monolayer TMDCs.

Charge-generation rates were calculated from the spatially resolved electromagnetic fields. Under optical excitation, the local absorption in the $MoS_2$ layer is given by the power dissipation density

$$P(\mathbf{r},\lambda) = \frac{1}{2}\omega\varepsilon_0\varepsilon_2(\mathbf{r},\lambda) \mid \mathbf{E}(\mathbf{r},\lambda) \mid^2 \quad (8)$$

The corresponding electron–hole pair generation rate is

$$G(\mathbf{r},\lambda) = \frac{P(\mathbf{r},\lambda)}{\hbar\omega} = \frac{1}{2}\frac{\omega\varepsilon_0\varepsilon_2|\mathbf{E}|^2}{\hbar\omega} \quad (9)$$

The generation maps were integrated over the MoS2 plane to obtain wavelength-dependent CGR spectra.

Radiative and non-radiative decay rates were evaluated by modeling the MoS2 exciton as an in-plane oscillating electric dipole located near the plasmonic cavity. In free space, the intrinsic dipole decay rate is given by:

$$\gamma_0 = \frac{\omega^3|p|^2}{3\pi\varepsilon_0\hbar c^3} \quad (10)$$

where $p$ is the dipole moment and $c$ is the speed of light. Within the cavity environment, the total decay rate was obtained from the total emitted power

$$\gamma_{tot} = \frac{P_{tot}}{P_0}\gamma_0 \quad (11)$$

where $P_0$is the dipole power in free space. Radiative emission was determined using far-field monitors,

$$\gamma_{rad} = \frac{P_{rad}}{P_0}\gamma_0 \quad (12)$$

while the non-radiative decay rate associated with metallic losses was obtained as:

$$\gamma_{nonrad} = \gamma_{tot} - \gamma_{rad}. \tag{13}$$

Near-field monitors were used to capture the full Poynting-vector flux surrounding the emitter,

$$P_{tot} = \oint \mathbf{S} \cdot dA \tag{14}$$

whereas far-field projection monitors were used to evaluate

$$P_{rad} = \oint_{far} \mathbf{S} \cdot dA. \tag{15}$$

This framework allows a clear separation of radiative enhancement and non-radiative plasmonic losses and enables quantitative evaluation of cavity-modified decay dynamics.

**Acknowledgments**

E.O.P. acknowledges financial support from the Scientific and Technological Research Council of Türkiye (TÜBİTAK) through Grant No. 222N308 within the CHIST-ERA program. Additional support was provided by the Turkish Academy of Sciences Outstanding Young Scientists Awards (GEBİP) 2025.

**Author Contributions**

A.E.Y. performed numerical simulations, carried out the calculations, and extracted the raw data underlying the reported results. A.E.Y. also contributed to the preparation of the manuscript. E.O.P. conceived the research idea, supervised the project, and led the preparation of the manuscript, including the development of the final figures and the main text. All authors discussed the results and contributed to the final version of the manuscript.

**Competing Interests**

The authors declare no competing financial or non-financial interests.

**Data Availability**

All data necessary to evaluate the conclusions of this work are included in the paper. Additional datasets supporting the findings of this study are available from the corresponding author upon reasonable request.